\documentclass{article}
\usepackage{mirex2010,cite}
\usepackage{graphicx}
\usepackage{amsmath}
\usepackage{amsfonts}
\usepackage{amssymb}
\usepackage{graphicx}
\usepackage{url}
\usepackage{multirow}
\usepackage{multicol}
\title{An end-to-end machine learning system for harmonic analysis of music}

\twoauthors
  {Yizhao Ni, Matt Mcvicar, and Tijl De Bie} {Intelligent Systems Lab\\ Department of Engineering Mathematics \\ University of Bristol\\ U. K.}
  {Raul Santos-Rodriguez} {Signal Theory and Communications Department\\ Universidad Carlos III de Madrid\\ Spain}

\begin{document}
\maketitle
\begin{abstract}
We present a new system for simultaneous estimation of keys, chords, and bass notes from music audio. It makes use of a novel chromagram representation of audio that takes perception of loudness into account. Furthermore, it is fully based on machine learning (instead of expert knowledge), such that it is potentially applicable to a wider range of genres as long as training data is available. As compared to other models, the proposed system is fast and memory efficient, while achieving state-of-the-art performance.
\end{abstract}
\section{Introduction}\label{sec:introduction}
Chords, along with the key and bassline, are essential mid-level features of western tonal music, and their evolution is fundamental to musical analysis.  In recent years, audio chord transcription and tonal key recognition have been very active fields \cite{labrosa_mirex_2010,MM_thesis,key_estimation_noland_2006,key_likelihood,probability_framework_key_chord_prediction_2007,unified_key_chord_prediction_2007,local_key_estimation_2009,rocher_key_estimation_2010} , and the increasing popularity of Music Information Retrieval (MIR) with applications using mid-level tonal features has established chord and key recognitions as useful and challenging tasks (see also e.g. the MIREX competitions).

Since chords and keys are musical attributes closely related to each other in western tonal music \cite{krumhansl_congnitive_of_musical_pitch_1990}, the idea to learn both progressions of a song simultaneously comes naturally. In general, such key/chord recognition systems are implemented using a HMM-like approach
, based on a set of features extracted from the audio signal
. A well-established audio feature for harmonic analysis is the \emph{chromagram} \cite{chroma_fujishima}. It is a $12$-dimensional representation of the harmonic content of the audio signal segmented into so-called \emph{frames}, and it reflects the distribution of energy along pitch classes. In this paper the chromagram for the audio signal $\mathbf{x}$ is denoted as $\bar{\mathbf{X}}\in \mathbb{R}^{12 \times T}$, with $T$ indicating the number of frames.

An HMM \cite{hmm_ref} commonly regards chromagrams and annotations as \emph{Observed} and \emph{Hidden} variables respectively. Let $\mathbf{k}\in \mathcal{A}_{k}^{1\times T}$ and $\mathbf{c}\in\mathcal{A}_{c}^{1\times T}$ be the key and the chord annotations of $\mathbf{x}$,
 where $\mathcal{A}_{k}$ and $\mathcal{A}_{c}$ represent the alphabets of keys and chords respectively. HMMs can then be used to formalize a probability distribution $P(\mathbf{k},\mathbf{c},\bar{\mathbf{X}}|\Theta)$ jointly for the chromagram feature vectors $\bar{\mathbf{X}}$ and the annotations, with $\Theta$ representing the parameters of this distribution. Given an HMM with optimal parameters $\Theta^{*}$, the key/chord recognition task is equivalent to finding $\{\mathbf{k}^{*},\mathbf{c}^{*}\}$ that maximize the joint probability
$
\{\mathbf{k}^{*},\mathbf{c}^{*}\}=\arg\max\limits_{\mathbf{k},\mathbf{c}}P(\mathbf{k},\mathbf{c},\bar{\mathbf{X}}|\Theta^{*}).
$

\begin{figure}[h!]
 \centerline{
 \includegraphics[scale=0.25]{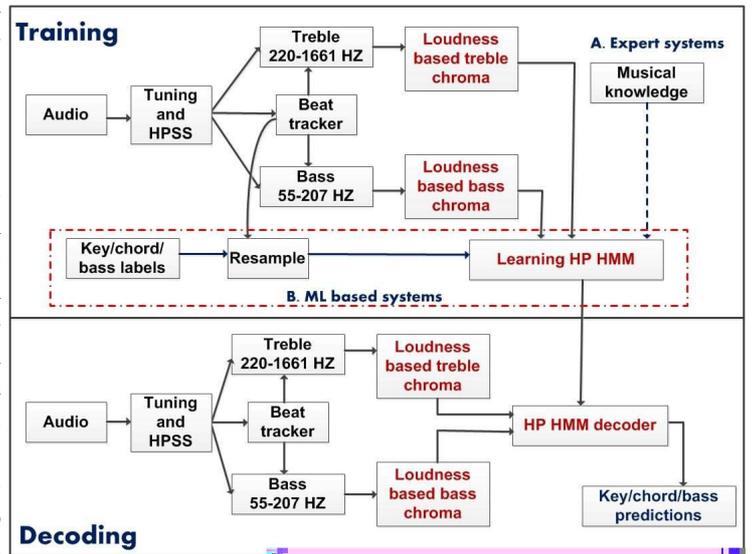}}
 \caption{The learning procedure (via Approach B)  of the proposed Harmony Progression (HP) system. The blocks in red show the novelties of the system.}
 \label{fig:KCB_process}
\end{figure}

\begin{figure}
 \centerline{
 \includegraphics[scale=0.23]{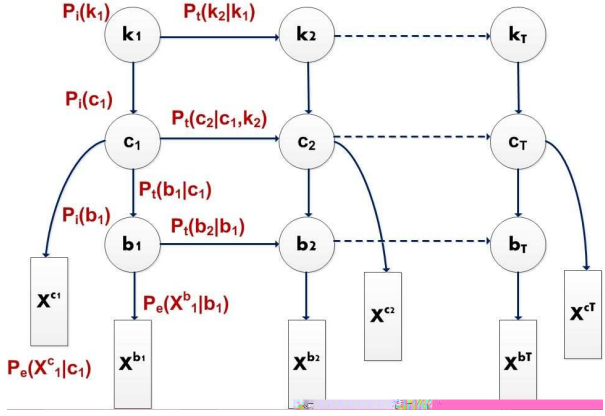}}
 \caption{The HMM topology of the HP system. The probabilities in red are parameters of the system, which are learnt via maximum likelihood estimation (MLE).}
 \label{fig:HMM_topology}
\end{figure}

Some existing key/chord recognition systems are based on Machine Learning (ML), where parameters are learned from a fully annotated training data set of features,
keys and chords: $\{\mathcal{X},\mathcal{K},\mathcal{C}\}=\{\bar{\mathbf{X}}^{n}\in\mathbb{R}^{12 \times T_{n}},\mathbf{k}^{n}\in \mathcal{A}_{k}^{1\times T_{n}},\mathbf{c}^{n}\in\mathcal{A}_{c}^{1\times T_{n}}\}_{n=1}^{N}$ (Approach B in Figure \ref{fig:KCB_process}) \cite{unified_key_chord_prediction_2007}.
However, most approaches are based at least partially on expert knowledge, where parameters are set on the basis of music theoretic knowledge of the developers (Approach A in Figure \ref{fig:KCB_process}) \cite{key_estimation_noland_2006,key_likelihood,probability_framework_key_chord_prediction_2007,local_key_estimation_2009,rocher_key_estimation_2010,MM_thesis}. For example, the key and chord transition parameters are set by hand, usually informed by perceptual key-to-key and chord-to-key relationships \cite{krumhansl_congnitive_of_musical_pitch_1990}. This contrasts with a clear tendency in Artificial Intelligence research to move away from systems based on expert knowledge to ML systems, e.g. in speech recognition, machine translation, computer vision, etc. 
We start from the premise that the key/chord recognition task is not different and propose the \emph{Harmony Progression (HP) system} for recognizing keys/chords from audio relying purely on ML techniques. The HP system is trained as illustrated in Figure \ref{fig:KCB_process} (Approach B) and the detailed HMM topology is depicted in Figure \ref{fig:HMM_topology}. Generally speaking, it is a simultaneous key/chord predictor that also identifies bass notes, going beyond most of the existing key/chord recognition systems \cite{key_estimation_noland_2006,key_likelihood,probability_framework_key_chord_prediction_2007,unified_key_chord_prediction_2007,local_key_estimation_2009,rocher_key_estimation_2010}. To our knowledge, the only system sharing a similar HMM topology is the expert knowledge based system  proposed in \cite{MM_thesis} -- the musical probabilistic model.

Compared with the MP system, the proposed HP system incorporates two additional major breakthroughs. Firstly, it utilizes a novel chromagram extraction method, supported with a well-founded physical interpretation. Secondly, 
our system is shown to be fast and memory-efficient in a case study. It also achieves an excellent tradeoff between performance and processing time in our experiments.

\section{System description}\label{sec:system_description}

\subsection{Loudness based chromagram}
Let $\mathbf{x}=[x_{1},\ldots,x_{T}]$ be an audio signal with $x_{t}$ indicating the sample data of the $t$-th frame, then the chromagram extraction assigns attributes (e.g.~power or amplitude) $\mathbf{X}\in \mathbb{R}^{S\times T}$ to a set of frequencies $F=\{f_{1},\ldots,f_{S}\}$ such that $\mathbf{X}$ reflects the energy distribution of the audio along these frequencies. In order to capture musically relevant information, the frequencies are selected from the equal-tempered scale, which may be tuned \cite{tuning} and vary between songs. Popular implementations of chromagram extraction are \emph{fixed bandwidth Fourier}  \cite{chroma_fujishima} and \emph{constant Q}  \cite{constantQ_1991} transforms.

The above two chromagram systems represent the salience of pitch classes in terms of a power or amplitude spectrum. We note however that perception of loudness is not linearly proportional to the power or amplitude spectrum, and
hence such chromagram representations do not accurately represent human perception of the audio's spectral content. Although there is an alternative chromagram that claimed to model human auditory sensitivity \cite{auditory_chromagram}, the proposed framework is very primitive. The chromagram still uses spectrum as pitch energy and it just utilizes an arc-tangent function to mimic pitch perception without any rigorous reference. In fact, the empirical study in \cite{loudness_definition_1933} showed that loudness is approximately linearly proportional to so-called \emph{sound power level}, defined as $\log_{10}$ of power spectrum. Therefore, we developed a novel \emph{loudness based chromagram}, which uses the  $\log_{10}$ scale of power spectrum. Mathematically, a sound power level (SPL) matrix is of the form
\begin{eqnarray*}
\mathcal{L}_{s,t}=10\log_{10}\left(\frac{\|X_{s,t}\|^{2}}{p_{ref}}\right),\textrm{ }s=1,\ldots,S, t=1,\ldots,T,
\end{eqnarray*}
where $p_{ref}$ indicates the fundamental reference power and
\begin{eqnarray*}
X_{s,t}=\sum\limits_{n=t-\frac{L_s}{2}}^{t+\frac{L_s}{2}}x_{n}w_{n}\exp \bigg( \frac{-2\pi st}{L_{s}} \bigg)
\end{eqnarray*}
is a constant Q transform with a frequency dependent bandwidth $L_{s}=Q\frac{SR}{f_{s}}$\footnote{$Q$ is a constant resolution fact which can be tuned by the cross-validation technique and $SR$ is the sampling rate of the audio signal.} and the hamming window $w_{n}$ \cite{constantQ_1991}.

Furthermore, low/high frequencies require higher sound power levels for the same perceived loudness as mid-fre\-quen\-cies \cite{loudness_definition_1933}. To compensate for this, we propose to use \emph{A-weighting} \cite{a_weighting_book_1999} to transform the SPL matrix into a representation of the perceived loudness of each of the pitches:
\begin{eqnarray*}
\mathcal{L}'_{s,t}=\mathcal{L}_{s,t}+A(f_{s}),\textrm{ }s=1,\ldots,S, t=1,\ldots,T,
\end{eqnarray*}
where
\begin{eqnarray*}
\begin{array}{l}
R_{A}(f_{s})=\frac{12200^2\cdot f_{s}^{4}}{(f_{s}^2+20.6^2)\cdot\sqrt{(f_{s}^2+107.7^2)(f_{s}^2+737.9^2)}\cdot (f_{s}^2+12200^2)}\\
A(f_{s})=2.0+20\log_{10}(R_{A}(f_{s})).\\
\end{array}
\end{eqnarray*}

It is known that loudnesses are additive if they are not close in frequency \cite{science_of_sound_1990}. This allows us to sum up loudness of sounds on the same pitch class, yielding:
\begin{equation*}
X'_{p,t}=\sum_{s=1}^{S}\delta(M(f_{s}),p)\mathcal{L}'_{s,t},\textrm{ }p=1,\ldots,12,t=1,\ldots,T.
\end{equation*}
Here $\delta$ denotes an indicator function and
$$M(f_{s})=\Bigg( \bigg\lfloor{12\log_{2}\left(\frac{f_{s}}{f_{A}}\right)+0.5}\bigg\rfloor+69\Bigg)\bmod 12$$
with $f_{A}$ denoting the reference frequency of the pitch $A4$ ($440$Hz in standard pitch). Finally, our loudness-based chromagram, denoted $\bar{X}_{p,t}$, is obtained by normalizing $X'_{p,t}$ using: 
\begin{equation*}
\bar{X}_{p,t}=\frac{X'_{p,t}-\min_{p'}X'_{p',t}}{\max_{p'}X'_{p',t}-\min_{p'}X'_{p',t}}.
\end{equation*}
Note that this normalization is invariant to the reference power and hence a specific $p_{ref}$ is not required.


\subsection{HP HMM topology}
The HP HMM topology consists of three hidden and two observed variables. The hidden variables correspond to the key $\mathcal{K}$, the chord $\mathcal{C}$ and the bass annotations $\mathcal{B}=\{\mathbf{b}^{n}\in \mathcal{A}_{b}^{1\times T_{n}}\}_{n=1}^{N}$. Under this representation, a chord is decomposed into two aspects: chord label and bass note. Take the chord \textrm{A:maj/3} for example, the chord state is $c=\textrm{A:maj}$ and the bass state is $b=\textrm{C\#}$. Accordingly, the observed chromagrams are decomposed into two parts: the treble chromagram $\bar{\mathbf{X}}^{\mathbf{c}}$ which is emitted by the chord sequence $\mathbf{c}$ and the bass chromagram $\bar{\mathbf{X}}^{\mathbf{b}}$ which is emitted by the bass sequence $\mathbf{b}$. The reason of applying this decomposition is that different chords can have the same bass note, resulting in similar chromagrams in low frequency domain.

Under this framework, the set $\Theta$ of a HP HMM has the following parameters
$$
\begin{array}{ll}
\Theta=&\big\{p_{i}(k_{1}),p_{i}(c_{1}),p_{i}(b_{1}),p_{t}(k_{t}|k_{t-1}),p_{t}(c_{t}|c_{t-1},k_{t}),\\
&p_{t}(b_{t}|c_{t}),p_{t}(b_{t}|b_{t-1}),p_{e}(\bar{\mathbf{X}}^{\mathbf{c}}_{t}|c_{t}),p_{e}(\bar{\mathbf{X}}^{\mathbf{b}}_{t}|b_{t})\big\},\\
\end{array}
$$
where $p_{i}$, $p_{t}$ and $p_{e}$ denote the initial, transition and emission probabilities respectively. The joint probability of the feature vectors $\{\bar{\mathbf{X}}^{\mathbf{c}},\bar{\mathbf{X}}^{\mathbf{b}}\}$ and the corresponding annotation sequences $\{\mathbf{k},\mathbf{c},\mathbf{b}\}$ of a song is then given by the formula\footnote{Note that we use $p_t(b_t|b_{t-1},c_t)=p_t(b_t|c_t)p_t(b_t|b_{t-1})$, which from a purely probabilistic perspective is not correct. However, this simplification reduces computational and statistical cost and results in better performance in practice.}
\begin{equation*}
\begin{array}{l}
P(\bar{\mathbf{X}}^{\mathbf{c}},\bar{\mathbf{X}}^{\mathbf{b}},\mathbf{k},\mathbf{c},\mathbf{b}|\Theta)=p_{i}(k_{1})p_{i}(c_{1})p_{i}(b_{1})\prod\limits_{t=2}^{T}p_{t}(k_{t}|k_{t-1})\\
p_{t}(c_{t}|c_{t-1},k_{t})p_{e}(\bar{\mathbf{X}}^{\mathbf{c}}_{t}|c_{t})p_{t}(b_{t}|c_{t})p_{t}(b_{t}|b_{t-1})p_{e}(\bar{\mathbf{X}}^{\mathbf{b}}_{t}|b_{t}).\\
\end{array}
\end{equation*}

The initial probabilities $p_{i}(\star)$
can be learnt via \textit{maximum likelihood estimation} (MLE). For example, $p_{i}(c)=\frac{\#(c_{1}=c)}{\#c_1}$ $\forall c\in\mathcal{A}_{c}$, where $\#$ indicates the number of.


For the transitions, $p_{t}(c|\bar{c},k)$ represents the probability of a chord change under a certain key. Since the chord transition is strongly influenced by the underlying key \cite{key_estimation_noland_2006}, this probability is modelled as key dependent. Under the assumption that relative chord transitions are key independent, we transposed all sequences to a common key $k$ and learn $p_{t}(c|\bar{c},k)$ from the transposed sequences. This allowed us to get 12 times as much information from the data source and the MLE solution is
$$
p_{t}(c|\bar{c},k)=\frac{\#(c_{t}=c \textrm{ \& } c_{t-1}=\bar{c} \textrm{ \& } k_{t}=k)}{\sum_{c'}\#(c_{t}=c' \textrm{ \& } c_{t-1}=\bar{c} \textrm{ \& } k_{t}=k)}, \forall c,\bar{c},k.
$$
Similarly, $p_{t}(k|\bar{k})$ is applied to model key changes during a song. $p_{t}(b|c)$ models the probability of a bass note under a chord label so as to capture chord inversions. A transition link $p_{t}(b|\bar{b})$ is also added, with the purpose of modelling the continuity of bass notes and capturing ascending and descending bassline progressions. These parameters are learnt via MLE, e.g.~$p_{t}(k|\bar{k})=\frac{\#(k_{t}=k\textrm{ }\&\textrm{ }k_{t-1}=\bar{k})}{\sum_{k'}\#(k_{t}=k'\textrm{ }\&\textrm{ }k_{t-1}=\bar{k})},\forall k,\bar{k}\in\mathcal{A}_{k}$.

Finally, emission probabilities $p_{e}(\bar{\mathbf{X}}^{\mathbf{c}}_{t}|c_{t})$ and $p_{e}(\bar{\mathbf{X}}^{\mathbf{b}}_{t}|b_{t})$ are modelled as $12$-dimensional Gaussians, of which the mean vectors and covariance matrices are learnt via MLE as well.

 \subsection{Search space reduction}
Given the optimal parameters $\Theta^{*}$ via MLE, the decoding task can be formalized as the computation of the key, chord and bass sequences $\{\mathbf{k}^{*},\mathbf{c}^{*},\mathbf{b}^{*}\}$ that maximize the joint probability
$
\{\mathbf{k}^{*},\mathbf{c}^{*},\mathbf{b}^{*}\}=\arg\max\limits_{\mathbf{k},\mathbf{c},\mathbf{b}}P(\bar{\mathbf{X}}^{\mathbf{c}},\bar{\mathbf{X}}^{\mathbf{b}},\mathbf{k},\mathbf{c},\mathbf{b}|\Theta^{*}).
$

This task can be solved using the \emph{Viterbi} algorithm \cite{hmm_ref}, whose computational complexity is $O\big(|\mathcal{A}_{k}|^2|\mathcal{A}_{c}|^2|\mathcal{A}_{b}|^2|T|\big)$. This is a huge search space, especially when one would like to use a large chord vocabulary \cite{MM_thesis}. In order to reduce the decoding time, we propose three constraints on the search space:

\subsubsection{Key transition constraint}
Music theory dictates that not all key changes are equally likely. If a song does change key, the modulation is most likely to move to a related key \cite{krumhansl_congnitive_of_musical_pitch_1990}. Thus, we suggest to rule out a priori the key transition that are seen the least often in the training set. Formally, this can be done by constraining the key transition probability as
$$p'_{t}(k|\bar{k})=\left\{
\begin{array}{ll}
p_{t}(k|\bar{k}) & \textrm{if }\#(k_{t}=k\textrm{ \& }k_{t-1}=\bar{k})>\gamma\\
0 & otherwise\\
\end{array}
\right.,
$$
where $\gamma$ is a positive integer indicating the threshold.

\subsubsection{Chord to bass transition constraint}
Similar to the key transition constraint, we can also constrain the chord to bass transitions. A constraint is imposed on $p_{t}(b|c)$ such that the bass notes can only be one of $\tau$ ($\tau \leq 12$) candidates for a given chord. The frequencies of each chord-to-bass emission are ranked and only the most common $\tau$ are permissible. Mathematically:
$$
p'_{t}(b|c)=
\begin{cases} p_{t}(b|c) & \text{if $b$ is one of the top $\tau$ bass notes for $c$}
\\
0 &\text{\emph{otherwise}}
\end{cases}.
$$
When $\tau=3$, the constraint is equivalent to using root position, first and second inversions of a chord.

\subsubsection{Chord alphabet constraint (CAC)}
It is unlikely that all chords will be used in a single song. Therefore, if it is possible to find out which chords are used in a song, we will be able to constrain the chord alphabet without loss of performance. One heuristic method is to utilize two-stage predictions. In particular, using a simple HMM with only chords as the hidden chain, we first apply a max-Gamma decoder \cite{hmm_ref}
to a song and obtain the most probable chords $\mathcal{A}'_{c}$. Then, we force the HP HMM chord transition probability to be zero for chords that are absent in this output:
$$
p'_{t}(c|\bar{c},k)=\left\{
\begin{array}{ll}
p_{t}(c|\bar{c},k) & \textrm{if }c,\bar{c}\in\mathcal{A}'_{c}\\
0 & otherwise\\
\end{array}
\right..
$$

\section{Experiments}\label{sec:experiment}

\subsection{Audio dataset and ground truth annotations}
The audio dataset used is the one used in the MIREX Chord Detection task 2010\footnote{\url{http://www.music-ir.org/mirex/wiki/2010:Audio_Chord_Estimation}}, which contains $217$ songs. The  ground truth key and chord annotations were obtained from \url{http://isophonics.net}, while the bass notes are extracted directly from the ground truth chord annotations.

\subsection{Preprocessing and chromagram feature extraction}
As shown in Figure \ref{fig:KCB_process}, we first converted our signals to mono $11025$ Hz, and separated the harmonic and percussive elements with the Harmonic/Percussive Signal Separation algorithm (HPSS) \cite{HPSS}. After tuning \cite{tuning} we computed loudness based chromagrams for each song. The frequency range of the bass chromagram was $A1$ to $G\sharp3$ ($55$Hz - $207.65$Hz), and that of the treble chromagram was $A3$ to $G\sharp6$ ($220$Hz - $1661.2$Hz). Finally, we estimated beat positions using the beat tracker presented in \cite{beat_tracker} and took the median chromagram feature between consecutive beats. We also beat synchronized our key/chord/bass annotations by taking the most prevalent labels between beats. The median feature vector with the corresponding beat-synchronized annotations is then regarded as one frame.

\subsection{Major/minor chord prediction}
In this experiment, we used a full key alphabet ($12$ major and $12$ minor keys), but restricted ourselves to a chord alphabet of $25$ chords ($12$ major, $12$ minor and no-chord). There were 13 bass states corresponding to the 12 pitch classes as well as a `no bass'. In accordance with the MIREX train-test setup, we randomly split $2/3$ of songs from each album to form the training set, while the remaining $1/3$ were used for testing. The same chord evaluation metric used in MIREX competition 2010 (denoted by `OR' and `WAOR'\footnote{'OR' refers to \emph{chord overlap ratio} in MIREX 2010 evaluation and `WAOR' refers to \emph{chord weighted average overlap ratio}.}) was applied to report chord prediction performance. Meanwhile, to evaluate the performance of key and bass predictions, the accuracy of predominant key prediction\footnote{Like in \cite{key_estimation_noland_2006,unified_key_chord_prediction_2007}, we regard the first key in the ground truth key sequence as the predominant key of this song, while the predicted predominant key will be the most prevalent key in the key prediction.} (denoted by `key-P') and the frame-based bass accuracy (denoted by `F-acc') were also reported. The experiment was repeated $102$ times to access variance. 


To compare chord and bass predictions, two HMM-Viterbi systems (denoted as HMM-C and HMM-B) are taken as baselines. For HMM-C, the observed variable is a concatenation of treble and bass chromagrams and the hidden states are $25$ chords; in HMM-B only bass chromagram is used as the observation and the hidden states are $13$ bass notes. Finally to compare key predictions, the performance of  a key-specific HMM \cite{unified_key_chord_prediction_2007} (denoted as K-HMM) is also reported.


\begin{table}[htb]
\begin{center}
\begin{tabular}{|c|cc|c|c|}
\hline
\multirow{2}{*}{System}&\multicolumn{2}{|c|}{Chord} & Key &Bass\\
\cline{2-5}
&OR [$\%$]&WAOR [$\%$]&key-P [$\%$]&F-acc [$\%$]\\
\hline
HMM-C&$77.82^{**}$&$77.22^{**}$&N/A&N/A\\
HMM-B&N/A&N/A&N/A&$73.62^{**}$\\
K-HMM&$78.22^{**}$ &$77.62^{**}$&$76.88^{*}$&N/A\\
HP&$\mathbf{79.37}$ &$\mathbf{78.82}$ &$\mathbf{77.36}$&$\mathbf{83.81}$\\
\hline
HP-P&$81.52$&$81.37$&$83.33$&$85.15$\\
\hline
\end{tabular}
%
%

\end{center} \caption{Performances for the baseline, key-specific HMM and HP systems on the major/minor chord prediction task. Bold numbers indicate the best results. The improvement of HP is significant at a level $<10^{-40}$ and $<10^{-1}$ over the performances marked by $^{**}$ and $^{*}$ respectively. The last line also shows the training set performance of HP.}\label{tab:major_minor_performance}
\end{table}

Table \ref{tab:major_minor_performance} shows the results and the significance of the improvement of the HP system over the other systems assessed using a paired t-test. The first row shows the results of the HMM-Viterbi chord prediction system using loudness based chromagram. This simple system already outperforms the best train-test system presented in MIREX 2010, whose results are $74.76\%$ (OR) and $73.37\%$ (WAOR)\footnote{The results are quoted from \url{http://nema.lis.illinois.edu/nema_out/mirex2010/results/ace/summary.html}.}, verifying the effectiveness of the novel loudness based chromagram extraction. Table \ref{tab:major_minor_performance} also indicates that increasing the complexity of models helps harmonic estimation, and that the HP system achieves the best performance on all evaluations.



To compare with the MIREX pre-trained systems, we trained and then tested our system on the whole dataset (denoted by HP-P). This provides an upper bound of performance the HP system can achieve, although of course is subject to overfitting the data. Compared with the best pre-trained system (namely MD1) presented in MIREX 2010, the results of which are $80.22\%$ (OR) and $79.45\%$ (WAOR), our pre-trained system achieves $>1\%$ improvement. Unfortunately we are unable to do a paired t-test on the results since we do not have their detailed prediction on each song.

\begin{figure}[h]
 \centerline{
 \includegraphics[scale=0.3]{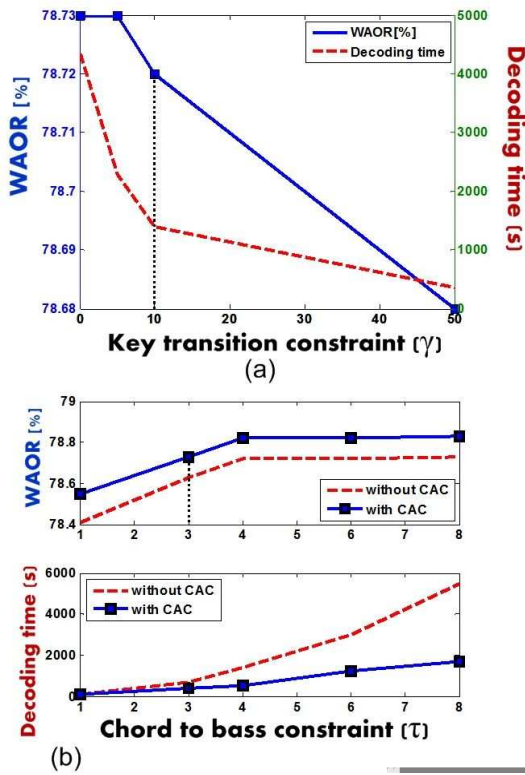}}
 \caption{
The performances and decoding times of HP using different search space reductions. The experiments in (a) were done without chord alphabet constraint and $\tau$ is fixed at $4$. In (b), `CAC' refers to chord alphabet constraint and the experiments were carried out with $\gamma$ fixed at $10$.}
 \label{fig:KCB_constraint}
\end{figure}

Finally we investigated the proposed search space reduction techniques. Figure \ref{fig:KCB_constraint} (a) shows that using a reasonable cutoff $\gamma$ can reduce the decoding time dramatically while retaining a high performance. The same trend is also observed when applying a reasonable $\tau$ to the chord to bass transition constraint (red dot curves in Figure \ref{fig:KCB_constraint} (b)). Furthermore, using a chord alphabet constraint (solid curves in Figure \ref{fig:KCB_constraint} (b)) did not decrease the performance (in fact it had a slight improvement), although the decoding time is also reduced. To summarize, by applying all these techniques, we are able to speed up decoding without decreasing the performance. Thanks to this, we can also apply HP to more complex chord representations in the next subsection.

\subsection{Full chord prediction}

Here we applied the proposed system to a chord recognition task using the chord dictionary used in \cite{MM_thesis}, with $12$ root notes and $11$ chord types\footnote{maj, min, maj/3, maj/5, maj6, maj7, min7, 7, dim, aug and `N'.}, resulting in $121$ unique chords. To the best of our knowledge current systems that can handle  this vocabulary are the musical probabilistic model (denoted by MP) \cite{MM_thesis} and Chordino \cite{nnls}.

We first compared the processing time and memory consumption of two songs\footnote{The information is quoted from \cite{MM_thesis} (page 78).} between our system and the state-of-the-art MP model (Table \ref{tab:mm_time_comparison}). Encouragingly, HP consumes less memory and is faster, even using a slower CPU.

\begin{table}[htb]
\begin{center}
\begin{tabular}{|c|cc|cc|}
\hline
\multirow{2}{*}{}&\multicolumn{2}{|c|}{Processing time (s)}&\multicolumn{2}{|c|}{Peak memory (G)}\\
\cline{2-5}
&HP&MP&HP&MP\\
\hline
Song 1&$58$&$131$&$0.48$&$6$\\
Song 2&$171$&$345$&$1.20$&$15$\\
\hline
\end{tabular}
\end{center} \caption{The comparison of processing time and memory consumption between the HP and MP systems. Song 1 is ``Ticket to Ride" ($190$s) and Song 2 is ``I Want You (She's So Heavy)" ($467$s). The MP results were performed on a computer running CentOS 5.3 with $8$ Xeon X5577 cores at $2.93$GHz, $24$G RAM. HP was run on a CentOS 5.6 computer with Intel (R) X5650 cores at $2.67$GHz, $24$G RAM.}\label{tab:mm_time_comparison}
\end{table}

Since MP is not publicly available, we instead compared HP to Chordino \cite{nnls} (denoted by CH) which uses the same NNLS chroma features as MP but a simpler model. Comparing with CH also seems more appropriate because its computation/memory cost is more reasonable and in line with HP.
For HP, the parameters $\tau$ and $\gamma$ are fixed at $3$ and $10$. All other parameters are trained using the whole dataset (denoted by HP-P). To assess generalization ability, we also computed the leave-one-out error for HP (denoted by HP-L). We used 3 performance metrics: chord precision (CP),
which scores $1$ if the ground truth and predicted chords are identical and $0$ otherwise (e.g.~the score between A:maj/3 and A:maj is $0$); note-based chord precision (NCP), which scores $1$ if all notes are identical between ground truth and predicted chords and $0$ otherwise (e.g.~the score between A:maj/3 and A:maj is $1$ but that between A:maj and A:maj7 is $0$), and the MIREX `WAOR' evaluation. All evaluations are performed with $1$ms sampling rate, as used in MIREX 2010 competition. Tests were done on a MAC with an Intel Duo Core $2.4$G CPU and $4$G RAM.

Table \ref{tab:mm_performance} shows a very large improvement over the baseline CH, even on the MIREX-style evaluation. Moreover, the full chord HP-P system achieves a further improvement on WAOR over the HP-P in the major/minor chord prediction task, again indicating that increasing the complexity of models helps harmonic estimation. Meanwhile, we found the cause of the low performance of CH is that it predicted many complex chords (notably $7th$s). This is a good strategy for the MIREX evaluation, that only measures the overlap recall between notes in predicted and ground truth chords. However, it does adversely affect the performances measured using CP and NCP.   Comparing the processing time, our system is slightly slower due to the separate calculation of bass and treble chromagrams. However, the decoding process is very fast and thus the system is still easy to apply to real world harmonic analysis tasks.


\begin{table}[htb]
\begin{center}
\begin{tabular}{|c|c|c|c|}
\hline
System& CP [$\%$]&NCP [$\%$]&WAOR [$\%$]\\
\hline
CH&$50.31$&$52.35$&$76.94$\\
HP-L&$63.63$&$65.24$&$81.05$\\
HP-P&$\mathbf{70.26}$&$\mathbf{71.96}$&$\mathbf{82.98}$\\
\hline
\end{tabular}

\vspace{0.1cm}

\begin{tabular}{|c|c|c|}
\hline
\multirow{2}{*}{System}&\multicolumn{2}{|c|}{Processing time (s)}\\
\cline{2-3}
&Feature extraction&Decoding \\
\hline
CH&\multicolumn{2}{|c|}{$\mathbf{9511}$}\\
\hline
HP&$12756$&$818$\\
\hline
\end{tabular}

\end{center} \caption{Performance (top) and processing time (bottom) for the baseline and HP systems on the full chord prediction task. Bold numbers refer to the best results. Note that for the CH system only the whole processing time is available.}\label{tab:mm_performance}
\end{table}

\section{Conclusions and future work}\label{sec:conclusion}
In this paper we propose a novel key, chord and bass simultaneous recognition system -- the HP system -- that purely relies on ML techniques. The experimental results verify that the HP system can achieve the state-of-the-art performance on chord recognition, and it can be sped up significantly using the search space reduction techniques without severely decreasing the performance.

HP uses a novel chromagram extraction method, which is inspired by loudness perception studies and achieves better recognition performance. Secondly, HP purely relies on ML techniques, which provides more flexibility in its applications and promises further improvements if more data becomes available. Finally, HP achieves an excellent tradeoff between performance and processing time, making it applicable to real world harmonic analysis tasks.

For future work, we aim to improve the processing time for chromagram extraction. This can be done by moving to faster programming languages such as C and C++. We will also move towards discriminative approaches using the same HMM topology, which might lead to a more robust and powerful harmonic analysis tool.

\bibliography{MIREX_2011_HP_toolbox}

\end{document}